\documentclass[prr,floatfix,reprint,amsmath,amssymb,superscriptaddress]{revtex4-1}
\usepackage{graphicx}
\usepackage{dcolumn}
\usepackage{bm}
\usepackage{color}

\newcommand{\BFS}{BaFe$_2$S$_3$}
\newcommand{\BFSE}{BaFe$_2$Se$_3$}
\newcommand{\Tstar}{$T^{*}$}
\newcommand{\TN}{$T_{\rm N}$}
\newcommand{\dzr}{$d_{3z^2 - r^2}$}
\newcommand{\dxsys}{$d_{x^2-y^2}$}

\newcommand{\dxz}{$d_{xz}$}

\newcommand{\dxy}{$d_{xy}$}

\begin{document}
%\preprint{APS/123-QED}
 %\linenumbers
\title{Dichotomy Between Orbital and Magnetic Nematic Instabilities in BaFe$_2$S$_3$}

\author{S. Hosoi}
\altaffiliation{These authors contributed equally to this work.}
\affiliation{Department of Advanced Materials Science, the University of Tokyo, Kashiwa, Chiba 277-8561, Japan}
\affiliation{Graduate school of Engineering Science, Osaka University, Toyonaka 560-8531, Japan}
\author{T. Aoyama}
\altaffiliation{These authors contributed equally to this work.}
\affiliation{Department of Physics, Graduate School of Science, Tohoku University, Sendai, Miyagi 980-8578, Japan}
\author{K. Ishida}
\affiliation{Department of Advanced Materials Science, the University of Tokyo, Kashiwa, Chiba 277-8561, Japan}
\author{Y. Mizukami}
\affiliation{Department of Advanced Materials Science, the University of Tokyo, Kashiwa, Chiba 277-8561, Japan}
\author{ K. Hashizume}
\affiliation{Department of Physics, Tohoku University, Sendai, Miyagi 980-8578, Japan}
\author{\\S. Imaizumi}
\affiliation{Department of Physics, Graduate School of Science, Tohoku University, Sendai, Miyagi 980-8578, Japan}
\author{Y. Imai}
\affiliation{Department of Physics, Graduate School of Science, Tohoku University, Sendai, Miyagi 980-8578, Japan}
\author{K. Ohgushi}
\affiliation{Department of Physics, Graduate School of Science, Tohoku University, Sendai, Miyagi 980-8578, Japan}
\author{Y. Nambu}
\affiliation{Institute for Materials Research, Tohoku University, Sendai, Miyagi 980-8577, Japan.}
\author{M. Kimata}
\affiliation{Institute for Materials Research, Tohoku University, Sendai, Miyagi 980-8577, Japan.}
\author{S. Kimura}
\affiliation{Institute for Materials Research, Tohoku University, Sendai, Miyagi 980-8577, Japan.}
\author{T. Shibauchi}
\affiliation{Department of Advanced Materials Science, the University of Tokyo, Kashiwa, Chiba 277-8561, Japan}

\begin{abstract}
Nematic orders emerge nearly universally in iron-based superconductors, but elucidating their origins is challenging because of intimate couplings between orbital and magnetic fluctuations. The iron-based ladder material BaFe$_2$S$_3$, which superconducts under pressure, exhibits antiferromagnetic order below $T_{\rm N}\sim 117$\,K and a weak resistivity anomaly at $T^*\sim 180$\,K, whose nature remains elusive. Here we report angle-resolved magnetoresistance (MR) and elastoresistance (ER) measurements in BaFe$_2$S$_3$, which reveal distinct changes at $T^*$. We find that MR anisotropy and ER nematic response are both suppressed near $T^*$, implying that an orbital order promoting isotropic electronic states is stabilized at $T^*$. Such an isotropic state below $T^*$ competes with the antiferromagnetic order, which is evidenced by the nonmonotonic temperature dependence of nematic fluctuations. In contrast to the cooperative nematic orders in spin and orbital channels in iron pnictides, the present competing orders can provide a new platform to identify the separate roles of orbital and magnetic fluctuations.
\end{abstract}

\maketitle
\section{INTRODUCTION}
The discovery of low-dimensional superconductivity in iron-based ladder compounds provides a new point of view in the study of iron-based superconductors \cite{HTakahashiNatMater2015}. Reduced dimensionality changes electronic structures dramatically, and the ground states of iron-based ladder materials show insulating properties in stark contrast to bad metal behaviors in the typical quasi-two dimensional BaFe$_2$As$_2$ (122) system. In spite of the totally different ground states, the ladder materials still have a stripe-type antiferromagnetic order similar to that of 122-system, suggesting the common physics between these two systems. More intriguingly, pressure-induced superconductivity emerges with the suppression of antiferromagnetism \cite{HTakahashiNatMater2015,JYingPRB2017}. These common features suggest unconventional pairing mechanisms of iron-based superconductivity robust against dimensionality.
 
The insulating nature of the ladder materials implies strong correlation effects due to the reduction of their dimensionality. Heavily hole-doped 122 system also has strong correlation effects since the nominal electron filling $3d^{5.5}$ in iron atoms, which is closer to half-filled states than $3d^6$ of non-doped 122-system, pushes towards the putative Mott insulating phase \cite{LDMediciPRL2014}. In contrast to the single-orbital case in e.g. cuprates, where only Hubbard interaction $U$ is dominant, in the multiorbital systems, Hund's coupling $J_{\rm H}$ plays an important role for increasing orbital-dependent correlation effects, leading to orbital selective Mott states \cite{LDMediciPRL2014,ZPYinNatMater2011}. An incoherent bad-metal conduction in these systems can be considered as a precursor phenomenon in the vicinity of Mott phases, and indeed some spectroscopy measurements reveal strongly orbital-dependent renormalized bands \cite{HDingJPCM2011,TYoshidaFP2014,MYiPRL2013}, although the system locates still far away from the parent Mott insulating state. The ladder materials show insulating behaviors, and thus the dimensionality may also serve as another promising parameter to tune the system to the Mott regime, which gives a new route to the study of orbital selective Mott phases.

%%%%%%%%%%%%%%%%%%FIG.1%%%%%%%%%%%%%%%%%%%
\begin{figure*}
	\includegraphics[width=\textwidth]{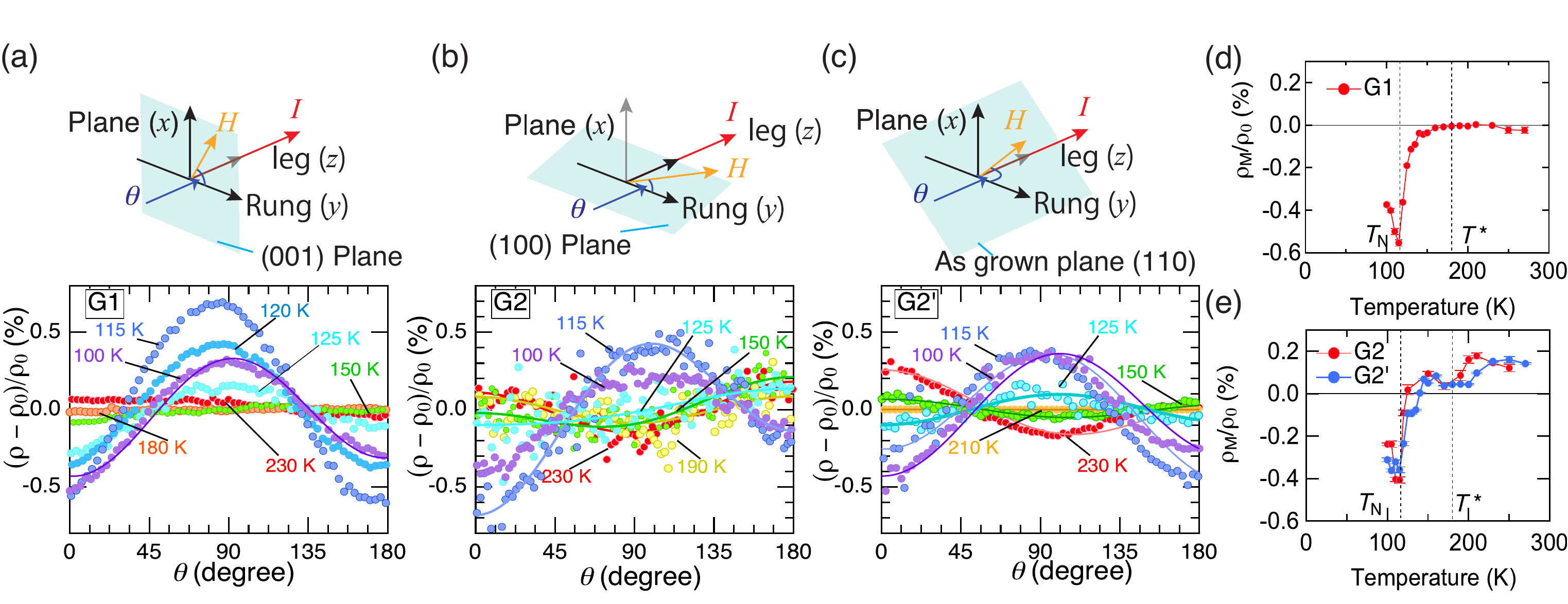}
	\protect\protect\caption{Angle-resolved magnetoresistance in \BFS. Three experimental geometries represent out-of-ladder, in-ladder, and as-grown plane anisotropy measurements as illustrated in the upper insets of (a-c) including our definitions of $x,y,z$ axes in this study and field angle $\theta$ with field rotation plane in each geometry. Angle dependence of MR can be fitted to Eq.\,(1) (solid lines are the fitting results). Temperature dependence of two-fold oscillation term is shown in (d) and (e) (broken lines are guides for the eyes) with error bar based on the standard deviation obtained from the fittings. The guide line for \Tstar, is determined from ER data as shown in Fig. \ref{fig:Elasto_all}(d).
		\label{fig:FAD} }
\end{figure*}
%%%%%%%%%%%%%%%%%%%%%%%%%%%%%%%%%%%%%%%

Superconductivity in \BFS\ under pressure occurs without significant crystal structure changes \cite{HTakahashiNatMater2015}, indicating that electronic state at ambient pressure smoothly connects to the superconductivity. Therefore, understanding the electronic states of \BFS\ at ambient pressure is fundamentally important to study the effects of dimensionality in iron-based materials. It has been established that \BFS\ exhibits the antiferromagnetic transition at $T_{\rm N}\sim 117$\,K with ferromagnetically aligned spins along rung directions \cite{HTakahashiNatMater2015}. One of the unsolved mysteries in \BFS\ is the so-called \Tstar\ anomaly  characterized by a weak bump-like features found in the temperature dependence of resistivity at $\sim180$\,K. The origin of the \Tstar\ anomaly possibly derives from an orbital-involved phase transition, as magnetization, neutron scattering, and muon spin relaxation measurements do not detect any magnetic signatures of phase transitions \cite{YHirataPRB,SChiPRL,LZhengPRBR2018}.  Orbital ordering has been intensively discussed in terms of electronic nematic orders in iron-based superconductors, as one of the promising candidates producing large in-plane anisotropy in the electronic state \cite{MYiPNAS2011}. Therefore, it is effective to evaluate anisotropy of the electronic state in order to verify the orbital order. Polarization dependence of  X-ray absorption spectroscopy (XAS) measurements have shown anisotropic electronic states, suggesting the existence of orbital order \cite{KTakuboPRB}. However, the observed polarization differences preserve up to 300\,K, and there is no characteristic change in spectrum around \Tstar. One possibility is that the \Tstar\ anomaly involves change of low-energy states outside the scope of the XAS measurements which generally probes valence bonds whose energy level is deep below the Fermi level. It is desirable to measure the anisotropy of transport properties that is very sensitive to low-energy quasiparticle excitations.

Here we report anisotropy of electronic states by measuring both field-angle resolved magnetoresistance (MR) and elastoresistance (ER) at zero field. From MR measurements, we can estimate the degree of anisotropy in the order parameter, and in contrast, ER couples to anisotropic electronic instability which corresponds to the nematic fluctuations. Measurements of these two physical values can provide complementary information on the rotational symmetry of the electronic states. The electronic properties of \BFS\ are known to be quite sensitive to the growth conditions \cite{XZhangSCPMA2018}. Here, by employing the method in Ref.\,\onlinecite{YHirataPRB}, we prepare single crystals without iron deficiency showing $T_{\rm N} \sim 117$\,K and $T^{*} \sim 180$\,K, which guarantee to acquire intrinsic information on the orbital order. Our measurements find the anisotropic state above \Tstar\ indicating the formation of leg-directed orbital ordering at high temperatures, consistent with XAS measurements. 
At around \Tstar, more isotropic electronic state forms, implying the formation of another orbital ordering (most likely \dxsys\ orbitals). 
Approaching $T_{\rm N}$ far below \Tstar, we find the enhancement of nematic fluctuations that are distinctly different from the high-temperature fluctuations, suggesting the presence of competing orbital and magnetic orders in this system.

\section{FIELD-ANGLE RESOLVED MAGNETORESISTANCE}

MR was measured up to $17.5$\,T and field angle dependence is resolved by using rotating probes in a superconducting magnet with three geometric configurations labeled by G1, G2 and G2', where magnetic field $H$ is rotated within out-of-ladder plane ($(001)$ or $xy$ plane), in-ladder plane ($(100)$ or $yz$ plane), and as-grown plane ($(110)$-leg plane), respectively. Hereafter, we define $z$ and $y$ axes along leg and rung directions, respectively, to discuss the electron $d$ orbitals [upper insets in Figs.\,\ref{fig:FAD}(a-c)]. From the view of space group $Cmcm$ of \BFS\cite{HHongJSSC1972,ZSGonen}, our introduced $x$, $y$, $z$ directions correspond to rung ($a = y$), plane ($b = x$), and leg ($c=z$), where $a$, $b$, $c$ are the crystal axes. In all cases, the direction of current $I$ is along the leg ($[001]$ or $z$) direction. 
Magnetic field angle dependence of resistivity can be fitted by 
\begin{equation}
 	\rho (\theta) = \rho_0 + \rho_{\rm M} \cos 2 \theta + \rho_{\rm H} \cos \theta
\end{equation}
where $\theta$ is the angle from the rung axis as shown in Figs.\,\ref{fig:FAD}(a-c) and its angle dependence of each term is determined through the restriction of orthorhombic symmetry \cite{RRBirss1963}.
Here, $\rho_{\rm H}$ represents the contribution from the Hall effect possibly caused by its strongly anisotropic structure. 
% which is smaller than $0.1$\,\% of angle-independent part $\rho_0$.
   The $\rho_{\rm M}$ term is the amplitude of angle dependence of MR and can be unambiguously separated from the Hall contribution because of the parity difference to magnetic field. The ratio of $\rho_{\rm M} / \rho_0$ represents the degree of field-angle dependence reflecting the anisotropy of the electronic states in the field-rotation planes.

The field-angle dependence of MR in G1 geometry starts to grow just above $T_{\rm N}$   and MR shows maximum (minimum) when magnetic field $H$ is applied perpendicular to the ladder plane (parallel to the rung direction). Magnetic moments aligned ferromagnetically 
along rung directions \cite{HTakahashiNatMater2015} can cause such angle dependence. The temperature dependence of the ratio $\rho_{\rm M} / \rho_0$ shows a sharp peak at $T_{\rm N}$ as shown in Fig.\,\ref{fig:FAD}(d). This suggests that antiferromagnetic fluctuations evolve towards $T_{\rm N}$, and such fluctuations start to develop below  $T^*$.
 
 %%%%%%%%%%%%%%Fig.2%%%%%%%%%%%%%%%%%%%%%%%%%%
 \begin{figure}
 	\includegraphics[width=\hsize]{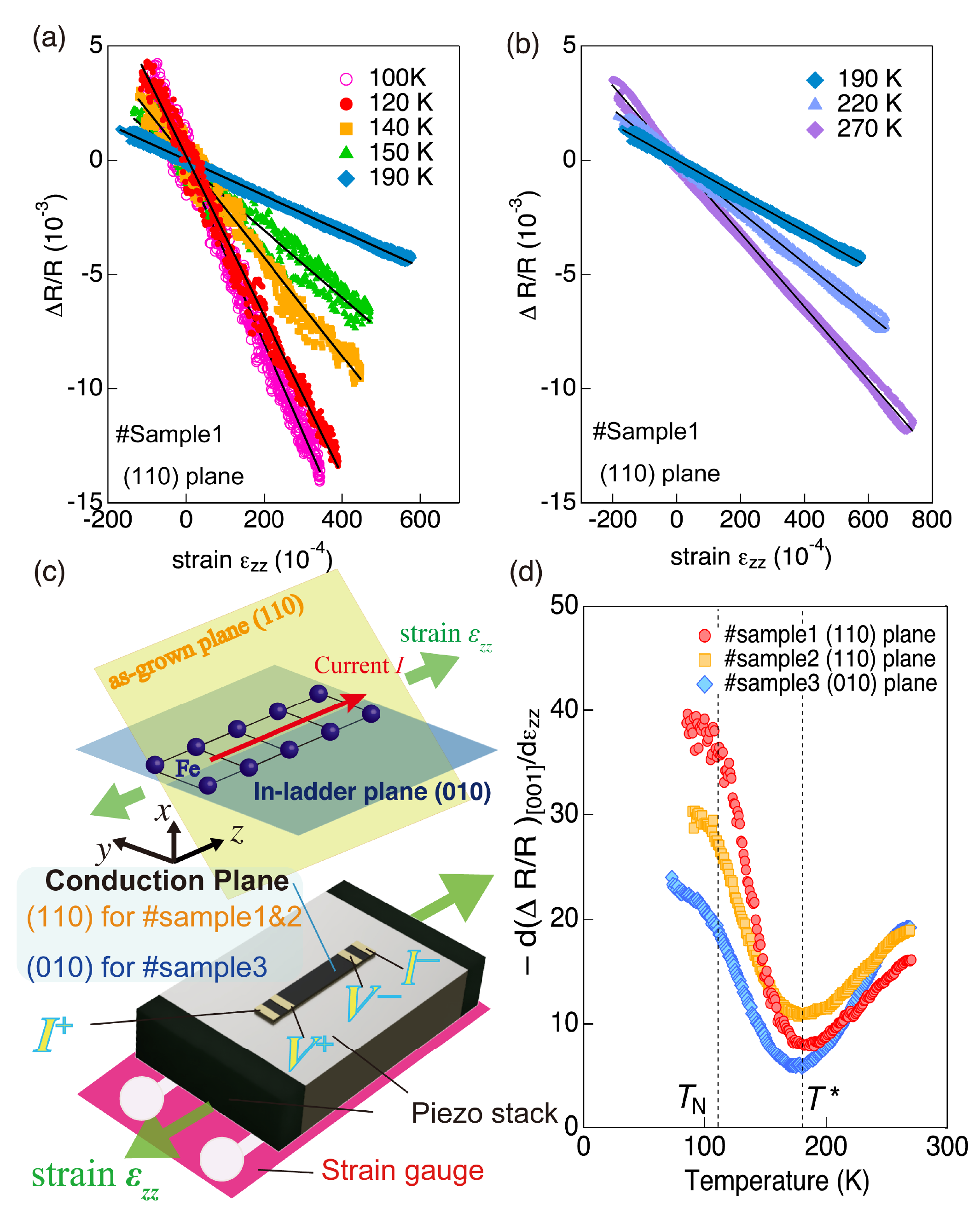}	
 	\protect\protect\caption{Strain dependence of resistance (a) below and (b) above \Tstar\ in \BFS\ (solid lines are fitting results). Here, strain is defined as $\varepsilon_{zz} =  \Delta L /L$, where $L$ is the sample length along the leg direction. (c) Experimental configuration of elastoresistance measurements: Both current and strain are applied along the $[001]$ leg direction. As-grown plane (110) is used as the conduction plane for Sample \#1 and \#2, which is tilted from in-ladder plane (010). In-ladder plane (010) is used for Sample \#3. Strain is evaluated by strain gauge glued on the backside of the piezostacks. (d) Temperature dependence of ER for three samples. \label{fig:Elasto_all} }
 \end{figure}
 %%%%%%%%%%%%%%%%%%%%%%%%%%%%%%%%%%%%%%%%%
 
To study in-plane anisotropy, G2 and G2' geometries are used, in which field rotation plane is parallel to and slightly tilted from the ladder plane, respectively, but both measurements reproduce essentially the same behavior as shown in Fig.\,\ref{fig:FAD}(e).  Below \TN, MR shows strong angle dependence similar to that of G1 geometry and takes a minimum when the $H$ is along the rung direction. At higher temperatures above \Tstar, however, $\rho_{\rm M}/\rho_0$ has the opposite sign and the magnitude of MR decreases with decreasing temperature and becomes very small around \Tstar. These results indicate that \BFS\ has an anisotropic electronic structure above \Tstar, which is distinct from that in antiferromagnetic state below \TN, and exhibits a phase transition around \Tstar\ leading to more isotropic electronic state.
%\subsection{Results}

\section{ELASTORESISTANCE}

ER, resistance change induced by uniaxial strain, is a powerful probe of nematic fluctuations \cite{JHChuScience2012,HHKuoScience2016}. We apply this technique to \BFS\ to study putative orbital ordering at $T^*$. The ER measurements require ideal geometry of thin bar shapes to control strain through the direct attachment of samples to piezoelectric device. Despite their fragile nature of single crystals, these requirements are achieved by cutting the samples with a wire-saw and polishing them carefully. In this study, both strain and current are applied along the $[001]$ leg direction [Fig.\,\ref{fig:Elasto_all}(c)]. We measured ER under two geometries with different conduction planes: the as-grown $(110)$ and ladder $(010)$ planes, the latter of which is a more appropriate configuration to probe the nature of ladder nematicity. In both cases, strain along the leg ($z$) direction most effectively changes electronic states in ladder plane, because strain along other directions would significantly alter the separation between ladders, which makes it more difficult to extract the ER response in the ladder. ER in tetragonal materials linearly couples to nematic fluctuations, but here the relationships between ER and nematicity fluctuations should be modified by the orthorhombic crystal structure of \BFS. Thus the Curie-Weiss analysis, which widely works for typical two-dimensional iron-based superconductors \cite{JHChuScience2012,HHKuoScience2016,SHosoiPNAS2016}, may not be valid in this case. However, as the leg distortions work as a conjugate field to ladder nematicity, strain-induced changes in resistance still couple in some form to nematic fluctuations even with distorted ladder structure.

Figures\,\ref{fig:Elasto_all}(a) and (b) show the strain response of resistance at low and high temperatures, respectively. The ER response is dominated by a linear slope with negative sign, in contrast to the metalization induced by hydrostatic pressure, which supports the successful control of anisotropic strain under little effect of symmetric strain. 
At high temperatures above \Tstar, the magnitude of ER slope decreases with decreasing temperatures. Below \Tstar, ER signal turns to grow and is enhanced towards $T_{\rm N}$. The temperature dependence of ER slope shown in Fig.\,\ref{fig:Elasto_all}(d) exhibits two anomalies; a kink around $T_{\rm N}\sim 117$\,K and a broad minimum around $T^*\sim 180$\,K. Here, three measured samples including different experimental geometries show similar behaviors, implying that our ER data successfully capture the essential features of nematic fluctuations.
The observed clear anomaly of ER provides strong evidence for the existence of electronic phase transition at \Tstar. In iron-based superconductors with Fe square lattice, ER nematic signal increases from high temperature side towards the nematic transition, but in the present case it shows an opposite decreasing trend. The behavior taking a minimum in ER can be explained by assuming multiple components of nematic fluctuations and competitive relationship between them.  

 %%%%%%%%%%%%%%Fig.3%%%%%%%%%%%%%%%%%%%%%%%%%%
\begin{figure*}
	\centering
	\includegraphics[width=0.8\textwidth]{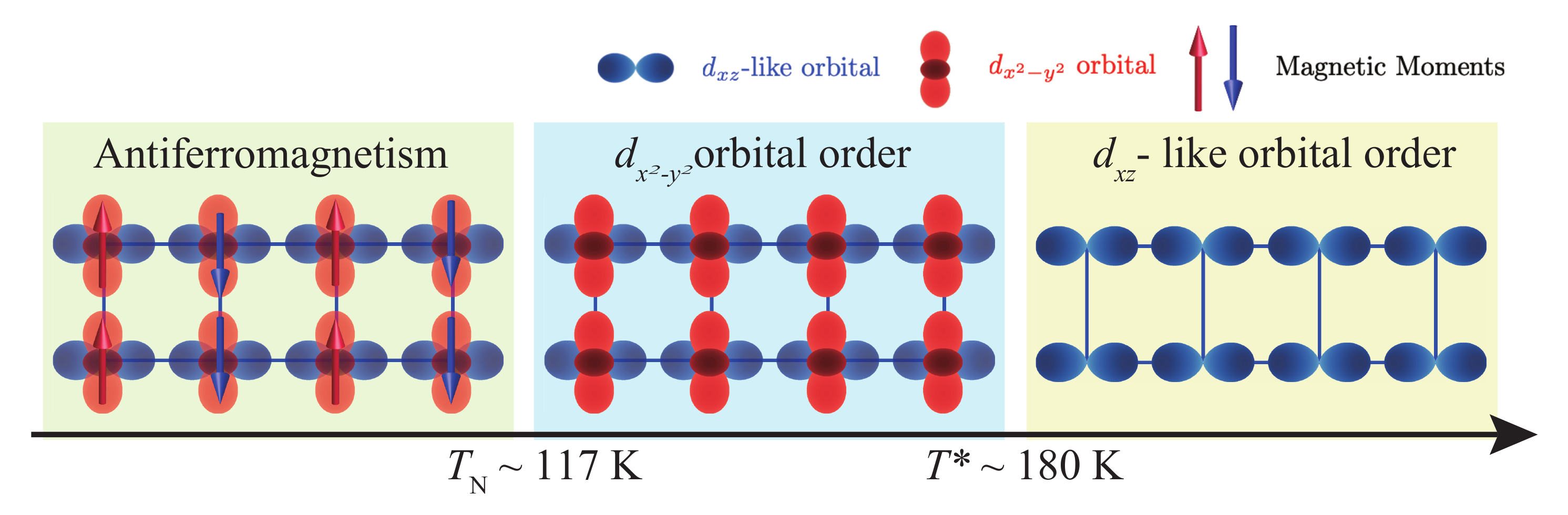}
	\protect\protect\caption{Schematic phase diagram of \BFS. Above \Tstar, leg-directed \dxz-like orbital order is stabilized and well-established stripe-type antiferromagnetic order exists below \TN. These two orders host anisotropic electronic structures and isotropic electronic states is accomplished in the gap region through the gradual orbital switching to \dxsys\ orbital.
		\label{fig:PD}}
\end{figure*}
%%%%%%%%%%%%%%Fig.3%%%%%%%%%%%%%%%%%%%%%%%%%%

\section{DISCUSSION}

Our measurements of field-angle dependent MR and ER consistently indicate a clear electronic change to a more isotropic state below \Tstar. Both measurements indicate that the origin of \Tstar\ is distinguished from antiferromagnetism below \TN\ and most likely involves orbital degrees of freedoms as previously speculated \cite{YHirataPRB,KTakuboPRB}. Although orbital order in iron-based superconductors has been discussed in the view of nematic order which induces in-plane anisotropy, the $T^*$ phase in \BFS\  rather reduces the anisotropy of electronic structure. 
 
Next we discuss possible orbital states above and below $T^*$. %Recent XAS measurements reveal anisotropy that links to orbital orders or fluctuations along leg directions even at room temperature \cite{KTakuboPRB}. There \dzr\ orbital ordering is labeled from the comparison of the two orbitals in the $e$\ state. 
Strong field-angle dependence of MR and large ER coefficient at $\sim 270$\,K is likely originating from leg-directed orbital ordering. %However, first-principle analyses suggested that \dzr\ orbitals have the peak of density of states, but it locates below the Fermi level, thus these orbitals are inactive in low energy state \cite{MTsuzukiPRB2015,RAritaPRB2015}. 
From the view of low-energy states, effective two-band model with the \dxsys\ orbitals and \dxz\ hybridized with \dxy\ orbitals have been proposed \cite{RAritaPRB2015,NDPatelPRB2016}. Since transport coefficients are generally sensitive to low-energy quasiparticle excitations, our transport anisotropy should be related to the imbalance between these two orbitals. Namely, the origin of the anisotropic state along the leg direction in high temperature region should come from \dxz-like orbital order in this model. Furthermore, the reduction of ER towards \Tstar\ from room temperature indicates a suppression of this leg-directed orbital fluctuations. This accomplishment of an isotropic electronic state can naturally be explained by the cancellation of anisotropy by the developments of another orbital \dxsys, which is elongated along rung direction and competes with the \dxz-like orbital. Here, the evolution of rung-directed order is gradual, so that it just compensates the anisotropy derived from leg-directed orbital order in range of $T_{\rm N} < T <T^*$, which is summarized in a schematic phase diagram in Fig.\,\ref{fig:PD}.  We note that if we assume \dzr\ orbital ordering in the place of \dxz-like orbitals as previously suggested from XAS \cite{KTakuboPRB}, discussion on gradual orbital switching to \dxsys\ at \Tstar\ is also applicable. These orbital decoupling behaviors emphasize the importance of the orbital selective Mott physics with Hund's coupling.

On the other hand, strong increase of ER below \Tstar\ can be understood by the contribution from magnetic nematicity associated with stripe-type antiferromagnetism, which also coincides with the evolution of MR. It is contrasting that orbital order becomes isotropic around \Tstar\ while the anisotropic antiferromagnetic order evolves.
 This is a quite unusual case because in most iron-based superconductors \dxz\ orbitals and stripe-type antiferromagnetism give a cooperative formation of nematic orders with the help of spin-orbit coupling. Another example that is distinct from the present case can be found in a related iron-based ladder compound \BFSE, which exhibits a structural distortion that supports the magnetic ordering \cite{JMCaronPRBR,YNambuPRB2012}. However, \dxsys\ orbital order hybridized along rung directions may support antiferromagnetic correlations along rungs, which is opposite to actual magnetic structure that magnetic moments align ferromagnetically along rungs \cite{HTakahashiNatMater2015}.
 We should also note that in \BFS\ magnetic ordering temperature \TN\ increases under pressure with a suppression of \Tstar\ \cite{TYamauchiPRL}, which is consistent with the competing nature of these two orders. Such an unusual competition of two orders in \BFS\ may provide hints to clarify the independent roles of spin and orbital degrees of freedoms.

%The schematic phase diagram of \BFS\ inferred from the present study is summarized in Fig.\,\ref{fig:PD}. \BFS\ has two different anisotropic states corresponding to magnetism below $T_{\rm N}$ and orbital ordering above \Tstar. An isotropic electronic state is found in the middle region, which has the competing nature with stripe-type antiferromagnetism. The origin of anisotropic orbital order at high temperatures has two possibilities: \dzr\ order suggested from XAS measurements or \dxz-like order based on first-principle analysis.  In either case, evolution of \dxsys\ orbital can cancel its anisotropy, leading to the formation of an isotropic state at \Tstar. Pinning down the order parameter above \Tstar\ or verifying orbital switching requires direct experimental determination of band structure, which deserves further studies.

The \Tstar\ anomaly is originally characterized by the slight improvement of conductivity \cite{YHirataPRB}. In fact, this can also be explained by the present orbital competition model discussed above: \dxz-like bands are strongly correlated and the evolution of \dxsys\ orbitals, which are more weakly correlated, leads to a decrease in resistivity. This resembles with incoherent-coherent crossover observed in hole-doped iron-based superconductors originating from the coexistence of localized and itinerant electrons due to orbital selective correlation effects \cite{FHardyPRL2013}.   
The density matrix renormalization group analysis on \BFS\ shows the orbital selective features in the perturbation response to hole doping: \dxz-like orbitals have a tendency to localize while \dxsys\ orbitals become coherent \cite{NDPatelPRB2016}. Experimentally, the coexistence of local and itinerant electrons is revealed by photoemission spectroscopy \cite{DOotsukiPRB2015}. The orbital switching from \dxz-like to \dxsys\ orbital proposed here based on the electronic anisotropy is thus consistent with the incoherent-coherent crossover associated with orbital selective Mottness.
 
%Finally, we discuss the effect of \dxy\ orbitals which have a significant contribution to \dxz\ orbitals suggested by first-principle calculations \cite{RAritaPRB2015}. In fact, hybridization bewtween \dxz\ and \dxy\ orbitals also push the system to more isotropic and coherent state, and thus consistently reproduce the present scenario. Namely, well-localized \dxz\ orbital provoke leg-directed anisotropic and incoherent states above \Tstar, and then the increase of contributions from \dxy\ orbitals via the hybridization compensates anisotropy and improves coherence, suggesting alternative description of \Tstar anomaly.

\section{CONCLUSION}
In summary, we performed field-angle resolved MR, and ER measurements in iron-ladder material BaFe$_2$S$_3$. Our results reveal strong anisotropy above \Tstar\ indicating leg-directed orbital ordering such as \dxz\ orbitals hybridized with $d_{xy}$, and the existence of an electronic phase transition leading to an isotropic state at \Tstar. A plausible microscopic origin of this phase transition is the orbital switching from $d_{xz}$-like orbitals to \dxsys\ coincided with incoherent-coherent crossover related to orbital selective Mott physics. Our proposed orbital order below \Tstar\ is an exotic state different from nematic orders in iron-based superconductors. One of the intriguing features is the dichotomy between orbital and magnetic nematic instabilities, which provides a new avenue for studying the roles of orbital and magnetic fluctuations.

\subsection*{ACKNOLEDGEMENTS}
We thank K. Takubo, T. Mizokawa, J. C. Palmstrom, T. Worasaran, and K. Izawa for helpful comments or fruitful discussions. We also thank  C. W. Hicks for technical supports on designing the wire-saw, and N. Abe, Y. Tokunaga, and T. Arima for experimental supports on X-ray diffraction measurements. This work was supported by Grants-in-Aid for Scientific Research (KAKENHI) (Grant Nos.\ JP15H02106, JP16H4007, JP16H04019, JP16K17732, JP17H05473, JP17H05474, JP17H06137, JP17J11382, JP18H01159, JP18H04302, JP18H05227, JP19H00649, and JP19H04683) and on Innovative Areas ``Quantum Liquid Crystals'' (Nos.\ JP19H05823, JP19H05824, and JP20H05162) from Japan Society for the Promotion of Science (JSPS). This work was partly performed at the High Field Laboratory for Superconducting Materials, Institute for Materials Research, Tohoku University (Project No.17H0202, 18H0204). X-ray diffraction measurements were partly performed using facilities of the Institute for Solid State Physics, the University of Tokyo.

\subsection{APPENDIX:SAMPLE PREPERATION}

\begin{figure}[h]
\centering
\includegraphics[width=\hsize]{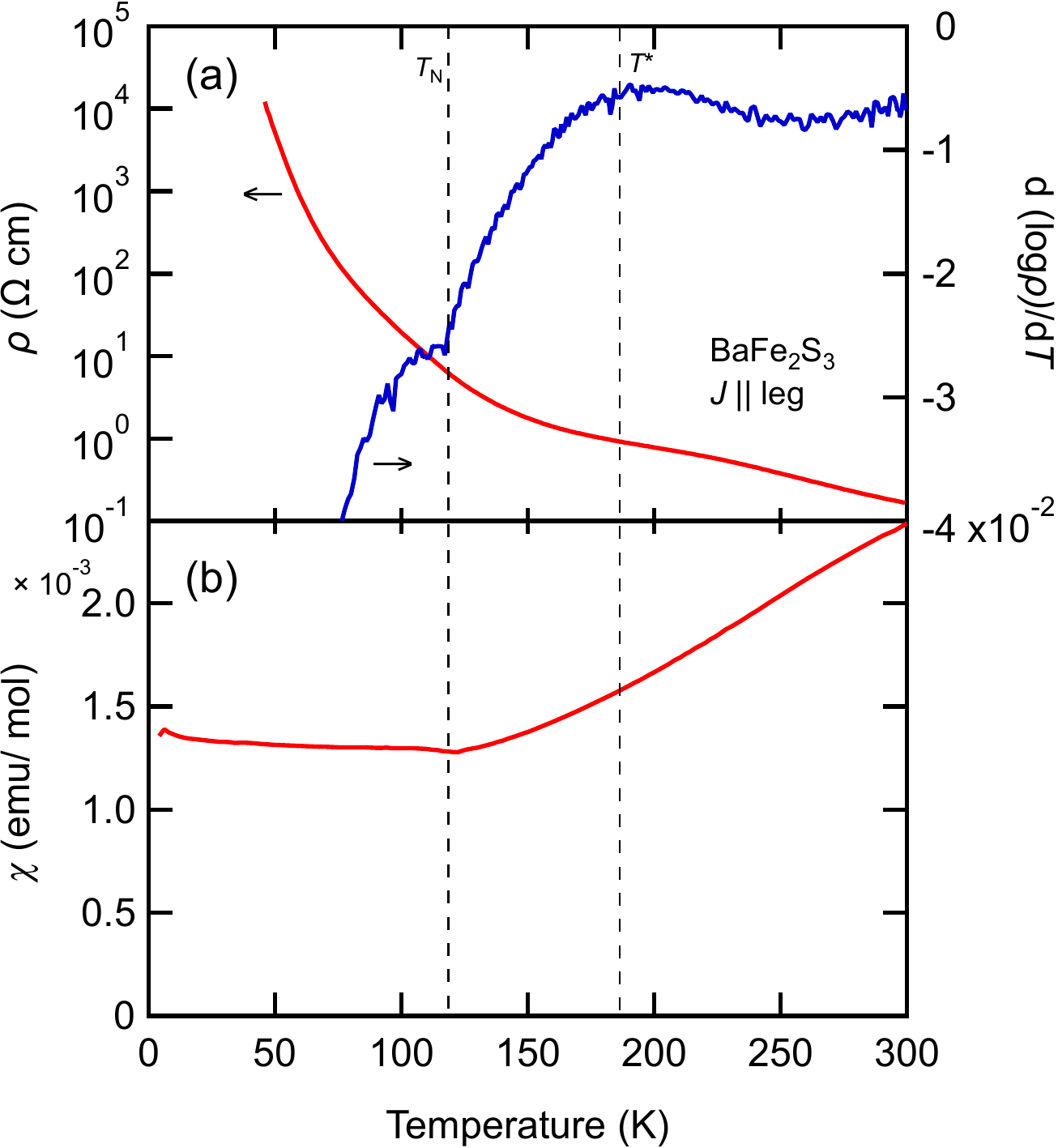}
\caption{
Temperature dependence of electronic properties of BaFe$_2$S$_3$. (a) Electrical resistivity ($\rho$) with the current applied parallel to the leg direction (left), and the temperature derivative of log$\rho$ (right). (b) Magnetic susceptibility ($\chi$) under the external magnetic field ($H$) of 1 T applied along the leg direction.
}
\label{rhoT}
\end{figure}

High-quality single crystals of BaFe$_2$S$_3$ were grown by the melt-growth method.
The starting materials are elemental Ba shots, Fe powders, and S powders with the ratio of 1 : 2.1 : 3.
The excess of iron is necessary to prevent the iron deficiency in the products as reported in Ref.~\cite{YHirataPRB}.
The starting materials in a carbon crucible were sealed into an evacuated quartz ampoule.
The ampoule was slowly heated up to 1373 K, kept for 48 hours, and slowly cooled to room temperature.
We characterized samples by measuring the temperature dependence of resistivity and magnetic susceptibility.
According to Hirata $et$ $al$.\cite{YHirataPRB}, the sample with the best quality has the highest antiferromagnetic transition temperature ($T_{\rm N}$) and the clearest anomaly in the resistivity at the possible orbital-order transition temperature ($T^*$).
Our electrical resistivity and magnetic susceptibility data shown in Fig. \ref{rhoT} exhibit clear anomalies at $T_{\rm N}$ = 117 K and $T^*$ = 180 K.
These features are consistent with those reported in Ref. \cite{YHirataPRB}.
We can therefore conclude that the sample used in this study is appropriate to acquire the intrinsic properties of BaFe$_2$S$_3$.

\subsection{APPENDIX:MAGNTETORESISTANCE UNDER THE MAGNETIC FIELD ALONG THE RUNG
DIRETION}

\begin{figure}[h]
\centering
\includegraphics[width=\hsize]{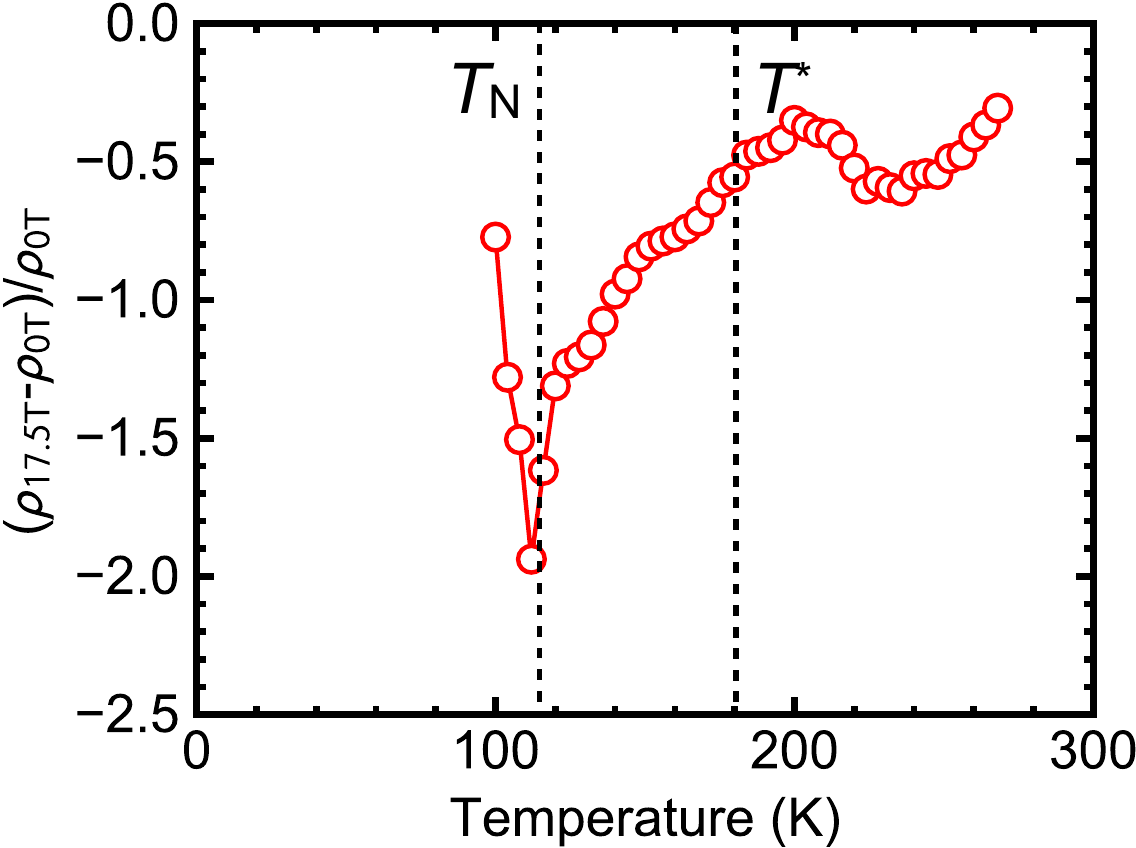}
\caption{
Temperature dependence of magnetoresistance for BaFe$_2$S$_3$  with the magnetic field of 17.5 T along the rung direction. The broken solid lines are guides for eyes. The guide for \Tstar\, is determined from ER data as shown in Fig.\ref{fig:Elasto_all}(d).
}
\label{MR_Tdep}
\end{figure}

In the main text, we deal with the magnetoresistance under the rotating magnetic field.
We here show the temperature dependence of magnetoresistance at fixed magnetic field direction along the rung direction (Fig. \ref{MR_Tdep}).
The data were obtained from the temperature dependence of resistivity ($\rho$) of in a constant magnetic field of 0 T and 17.5 T.
One can clearly see the suppressed magnetoresistance with a bump-like structure at around $T^{*}$  and enhanced negative magnetoresistance at $T_{\rm N}$, although the $T^{*}$ position estimated from ER data is slightly deviated from the top of the bump structure.
The former and the latter correspond to the formation of the isotropic electronic states and the enhancement of magnetic fluctuations, respectively.
The details of each transitions are discussed in the main text.

\begin{thebibliography}{99}
	\bibitem{HTakahashiNatMater2015}H. Takahashi, A. Sugimoto, Y. Nambu, T. Yamauchi, Y. Hirata, T. Kawakami, M. Avdeev, K. Matsubayashi, F. Du, C. Kawashima, H. Soeda, S. Nakano, Y. Uwatoko, Y. Ueda, T. J. Sato, and K. Ohgushi,
	 Pressure-induced superconductivity in the iron-based ladder material \BFS.
	  Nat. Mater. {\bf 14}, 1008 (2015).
	\bibitem{JYingPRB2017}J. Ying, H. Lei, C. Petrovic, Y. Xiao, and V.-V. Struzhkin, Phys. Rev. B {\bf 95}, 241109(R) (2017).
	\bibitem{LDMediciPRL2014}L. de' Medici, G. Giovannetti, and M. Capone,
	Interplay of magnetism and superconductivity in the compressed Fe-ladder compound BaFe$_2$Se$_3$.
	 Phys. Rev. Lett. {\bf 112}, 177001 (2014).
	\bibitem{ZPYinNatMater2011}Z. P. Yin, K. Haule, and G. Kotliar,
	Kinetic frustration and the nature of the magnetic and paramagnetic states in iron pnictides and iron chalcogenides.
	 Nat. Mater. {\bf 10}, 932 (2011).
	\bibitem{HDingJPCM2011}H. Ding, K. Nakayama, P. Richard, S. Souma, T. Sato,
	T. Takahashi, M Neupane, Y.-M. Xu, Z.-H. Pan, A. V. Fedorov, Z.Wang, X.Dai, Z. Fang, G. F. Chen, J. L. Luo, and N. L. Wang,
	Electronic structure of optimally doped pnictide Ba$_{0.6}$K$_{0.4}$Fe$_2$As$_2$: a comprehensive angle-resolved photoemission spectroscopy investigation.
	 J. Phys.: Condens. Matter {\bf 23}, 135701 (2011).
	\bibitem{TYoshidaFP2014}T. Yoshida, S. Ideta, I. Nishi, A. Fujimori, M. Yi, R. G. Moore, S. K. Mo, D. -H. Lu, Z.-X. Shen, Z. Hussain, K. Kihou, P. M. Shirage, H. Kito, C.-H. Lee, A. Iyo, H. Eisaki, and H. Harima,
	Orbital character and electron correlation effects on two- and three-dimensional Fermi surfaces in KFe$_2$As$_2$ revealed by angle-resolved photoemission spectroscopy.
	 Front. Phys. {\bf 2}, 17 (2014).
	\bibitem{MYiPRL2013}M. Yi, D. H. Lu, R. Yu, S. C. Riggs, J.-H. Chu, B. Lv, Z. K. Liu, M. Lu, Y.-T. Cui, M. Hashimoto, S.-K. Mo, Z. Hussain, C. W. Chu, I. R. Fisher, Q. Si, and Z.-X. Shen,
	Observation of Temperature-Induced Crossover to an Orbital-Selective Mott Phase
	 in $A_x$Fe$_{2-y}$
	 Se$_2$ ($A$=K, Rb) Superconductors.
	 Phys. Rev. Lett. {\bf 110}, 067003 (2013).
	\bibitem{YHirataPRB}Y. Hirata, S. Maki, J.-I. Yamaura, T. Yamauchi, and K. Ohgushi,
	Effects of stoichiometry and substitution in quasi-one-dimensional iron chalcogenide \BFS.
	 Phys. Rev. B {\bf 92}, 205109 (2015).
	\bibitem{SChiPRL}S. Chi, Y. Uwatoko, H. Cao, Y. Hirata, K. Hashizume, T. Aoyama,  and K. Ohgushi,
	Magnetic Precursor of the Pressure-Induced Superconductivity in Fe-Ladder Compounds.
	 Phys. Rev. Lett. {\bf 117}, 047003 (2016).
	\bibitem{LZhengPRBR2018}L. Zheng, B. A. Frandsen, C. Wu, M. Yi, S. Wu, Q. Huang, E. Bourret-Courchesne, G. Simutis, R. Khasanov, D.-X. Yao, M. Wang, and R. J. Birgeneau,
	Gradual enhancement of stripe-type antiferromagnetism in the spin-ladder material \BFS\, under pressure.
	 Phys. Rev. B {\bf 98}, 180402(R) (2018).
	\bibitem{MYiPNAS2011}M. Yi, D. Lu, J.-H. Chu, J. G. Analytis, A. P. Sorini, A. F. Kemper, B. Moritz, S.-K. Mo, G. Moore, M. Hashimoto, W.-S. Lee, Z. Hussain, T. P. Devereaux, I. R. Fishser, and Z.-Xun. Shen,
	Symmetry-breaking orbital anisotropy observed for detwinned Ba(Fe$_{1-x}$Co$_x$)$_2$As$_2$ above the spin density wave transition.
	 Proc. Natl. Acad. Sci. USA {\bf 108}, 6878 (2011).
	\bibitem{KTakuboPRB}K. Takubo, Y. Yokoyama, H. Wadati, S. Iwasaki, T. Mizokawa, T. Boyko, R. Sutarto, F. He, K. Hashizume, S. Imaizumi, T. Aoyama, Y. Imai, and K. Ohgushi,
	Orbital order and fluctuations in the two-leg ladder materials BaFe$_2$X$_3$ ($X$ = S and Se) and CsFe$_2$Se$_3$.
	 Phys. Rev. B {\bf 96}, 115157 (2017).
	\bibitem{XZhangSCPMA2018}X. Zhang, H. Zhang, Y. H. Ma, L. L. Wa, J. N. Chu, T. Hu, G. Mu, Y. M. Lu, C. B. Cai, F. Q. Huang, and X. M. Xie,
	{\it In situ} aneealing effects on magnetic properties and variable-range hopping of iron-based ladder material \BFS.
	 Sci. China Phys. Mech. Astron, {\bf61}, 77421 (2018).
	\bibitem{TYamauchiPRL}T. Yamauchi, Y. Hirata, Y. Ueda, and K. Ohgushi,
	Pressure-Induced Mott Transition Followed by a 24-K Superconducting Phase in \BFS.
	 Phys. Rev. Lett. {\bf 115}, 246402 (2015).
	\bibitem{HHongJSSC1972}H. Hong and H. Steinfin,
	The Crystal Chemistry of Phases in the Ba-Fe-S and Se Systems.
	 J. Solid State Chem. {\bf5}, 93 (1972).
	\bibitem{ZSGonen}Z. S. G${}\rm \ddot{o}$nen, P. Fournier, V. Smolyaninova, R. Greene, F. M. Araujo-Moreira, and B. Eichhorn,
	Magnetic and Transport Properties of Ba$_6$Fe$_8$S$_{15}$ and BaFe$_2$S$_3$
	: Magnetoresistance 
	in a Spin-Glass-Like Fe(II) System.
	 Chem. Mater., {\bf12}, 3331 (2000).
	\bibitem{RRBirss1963}R. R. Birss,
	Macroscopic symmetry in space-time.
	 Rep. Prog. Phys.  {\bf 26}, 307 (1963).
	\bibitem{JHChuScience2012}J.-H. Chu, H.-H. Kuo, J. G. Analytis, and I. R. Fisher,
	Divergent Nematic Susceptibility in an Iron Arsenide Superconductor.
	 Science {\bf337}, 710 (2012).
	\bibitem{HHKuoScience2016}H.-H. Kuo, J.-H. Chu, J. C. Palmstrom, S. A. Kivelson, and I. R. Fisher,
	Ubiquitous signatures of nematic quantum criticality in optimally doped Fe-based superconductors.
	 Science {\bf352}, 958 (2016).
	\bibitem{SHosoiPNAS2016}S. Hosoi, K. Matsuura, K. Ishida, H. Wang, Y. Mizukami, T. Watashige, S. Kasahara, Y. Matsuda, and T. Shibauchi,
	Nematic quantum critical point without magnetism in FeSe$_{1-x}$S$_x$ superconductors.
	 Proc. Natl. Acad. Sci. {\bf 113}, 8139 (2016).
%	\bibitem{MTsuzukiPRB2015}M.-T. Suzuki, R. Arita, and H. Ikeda, Phys. Rev. B {\bf 92}, 085116 (2015).
	\bibitem{RAritaPRB2015}R. Arita, H. Ikeda, S. Sakai, and M.-T. Suzuki,
	{\it Ab initio} downfolding study of the iron-based ladder superconductor \BFS. 
	 Phys. Rev. B {\bf 92}, 054515 (2015).
	\bibitem{NDPatelPRB2016}N. D. Patel, A. Nocera, G. Alvarez, R. Arita, A. Moreo, and E. Dagotto,
	Magnetic properties and pairing tendencies of the iron-based superconducting ladder 
BaFe$_2$S$_3$: Combined {\it ab initio} and density matrix renormalization group study.
	 Phys. Rev B {\bf 94}, 075119 (2016). 
	\bibitem{JMCaronPRBR}J. M. Caron, J. R. Neilson, D. C. Miller, A. Llobet, and T. M. McQueen,
	Iron displacements and magnetoelastic coupling in the antiferromagnetic spin-ladder compound BaFe$_2$Se$_3$.
	 Phys. Rev. B {\bf 84}, 180409(R) (2011).
	\bibitem{YNambuPRB2012}Y. Nambu, K. Ohgushi, S. Suzuki, F. Du, M. Avdeev, Y. Uwatoko, K. Munakata, H. Fukazawa, S. Chi, Y. Ueda, and T. J. Sato,
	Block magnetism coupled with local distortion in the iron-based spin-ladder compound BaFe$_2$Se$_3$.
	 Phys. Rev B {\bf 85}, 064413 (2012).
	\bibitem{FHardyPRL2013}F. Hardy, A. E. B${}\rm \ddot{o}$hmer, D. Aoki, P. Burger, T. Wolf, P. Schweiss, R. Heid, P. Adelmenn, Y. X. Yao, G. Kotliar, J. Schmailian, and C.  Meingast,
	Evidence of Strong Correlations and Coherence-Incoherence Crossover in the Iron Pnictide Superconductor KFe$_2$As$_2$.
	 Phys. Rev. Lett. {\bf 111}, 027002 (2013).
	\bibitem{DOotsukiPRB2015}D. Ootsuki, N. L. Saini, F. Du, Y. Hirata, K. Ohgushi, Y. Ueda, and T. Mizokawa,
	Coexistence of localized and itinerant electrons in BaFe$_2X_3$ ($X$=S and Se) revealed by photoemission spectroscopy
	 Phys. Rev. B {\bf 91}, 014505 (2015).   
\end{thebibliography}
\end{document}